\documentclass[aps,prb,showpacs,twocolumn,superscriptaddress]{revtex4}
\usepackage{bm,color}
\usepackage{graphicx}
\usepackage{amsmath}
\usepackage{color}
\pdfoutput=1

\begin{document}
\title {Creation of nanometric magnetic skyrmions by global application of circularly polarized microwave magnetic field}
\author{Masayuki Miyake}
\author{Masahito Mochizuki}
\affiliation{Department of Applied Physics, Waseda University, Okubo, Shinjuku-ku, Tokyo 169-8555, Japan}
\begin{abstract}
From a theoretical perspective, we demonstrate that nanometric magnetic skyrmions are created by application of a circularly polarized microwave magnetic field to a thin-plate Dzyaloshinskii-Moriya ferromagnet with fabricated rectangular holes. This phenomenon is caused by an effective steady magnetic field perpendicular to the microwave-polarization plane induced by the rotating magnetic field and the intense interference of spin waves excited by this magnetic field due to the hole structure, which causes reversals of local magnetizations and results in the formation of skyrmions. Our proposal provides a new option to write or create magnetic textures the sizes of which are much smaller than the spot size of the external stimulus such as magnetic field, light, and microwave.
\end{abstract}
\maketitle

\section{Introduction}
\begin{figure}[t]
\includegraphics[width=1.0\columnwidth]{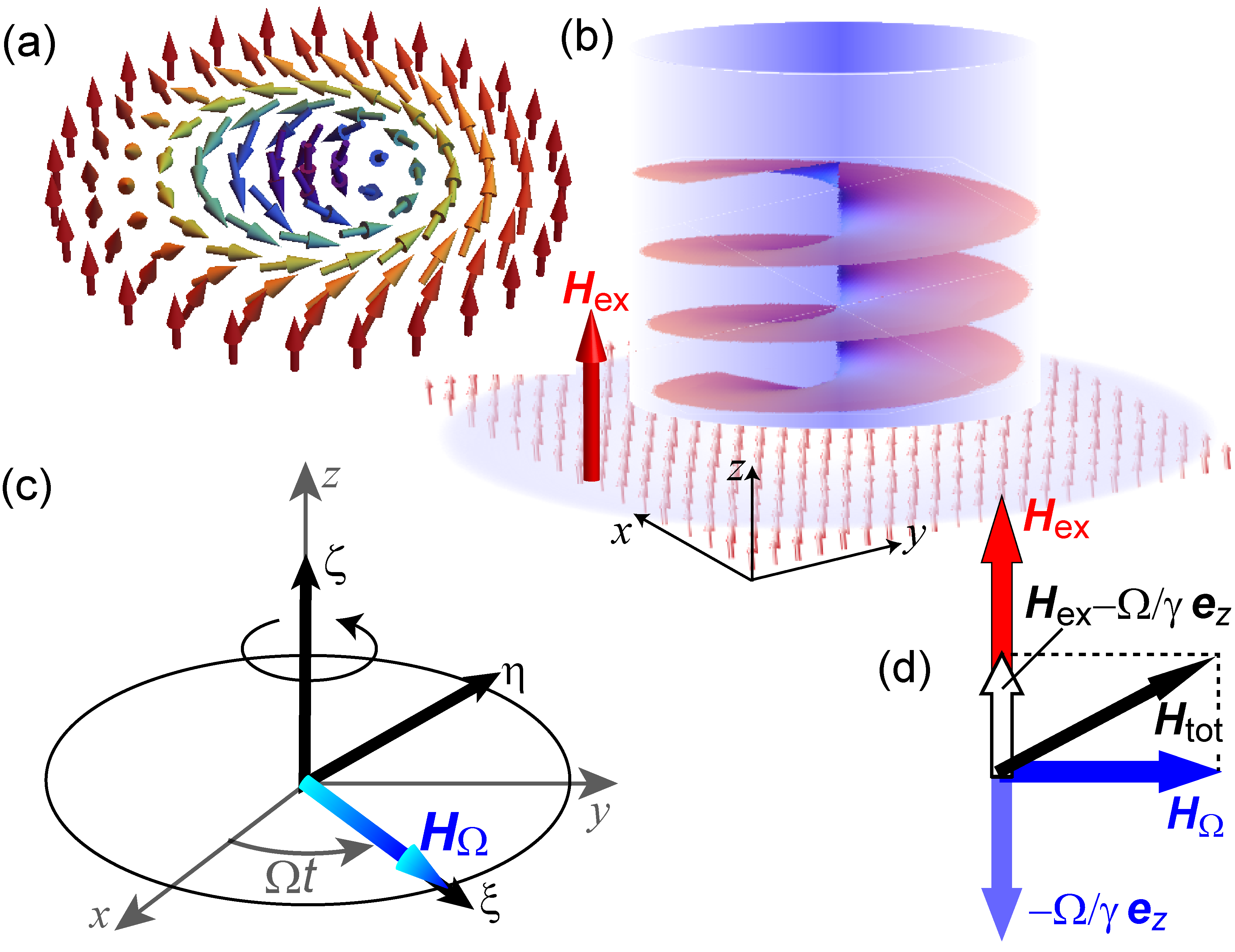}
\caption{(a) Magnetization configuration of a magnetic skyrmion, (b) Schematic of the application of a circularly polarized microwave magnetic field to a ferromagnetic state in a thin-plate sample of a chiral-lattice magnet. (c) Rest coordinates ($x$, $y$, $z$) and rotating coordinates ($\xi$, $\eta$, $\zeta$) with $\xi$ axis parallel to the circulating microwave magnetic field $\bm H_\Omega=H_\Omega(\cos \Omega t, \sin \Omega t, 0)$ in the $xyz$ representation. (d)  Effective magnetic field $\bm H_{\rm tot}$ acting on magnetizations where its in-plane component comes from the rotating microwave magnetic field $\bm H_\Omega$, whereas its out-of-plane component is composed of two contributions, that is, the static external magnetic field $\bm H_{\rm ex}$ and an effective steady component $-\Omega/\gamma$$\bm e_z$ induced by the circularly polarized microwave.}
\label{Fig01}
\end{figure}
In ferromagnets with broken spatial inversion symmetry, the Dzyaloshinskii-Moriya (DM) interaction, which favors a rotating magnetization alignment with 90$^\circ$ pitch angles, becomes active and competes with the ferromagnetic-exchange interaction, which favors a parallel magnetization alignment. This keen competition results in the formation of magnetic skyrmions, that is, particle-like magnetic textures with a quantized topological invariant under an external magnetic field of appropriate strength [see Fig.~\ref{Fig01}(a)]~\cite{Seki15,Everschor18,Nagaosa13,Fert13}. Skyrmions belong to a different topological class from the ferromagnetic state and thus cannot be created or erased in the ferromagnetic state under continuous deformation of the magnetization alignment, which gives their robustness and stability against external disturbances. In addition, magnetic skyrmions can be driven at ultralow energy costs. For example, it is known that their translational motion can be induced by the injection of spin-polarized electric currents via the spin-transfer torque mechanism; its threshold current density turns out to be five or six orders of magnitude smaller than that required to drive other magnetic textures~\cite{Jonietz10,YuXZ12,Iwasaki13a}. For this reason, the skyrmion has attracted a great deal of research interest in the spintronics field from the viewpoint of their potential applications to high-performance magnetic memory devices~\cite{Fert13,Tomasello14,Koshibae15}.

Many experimental demonstrations and theoretical proposals of skyrmion creation have been reported so far~\cite{Kang16}, in which several kinds of external stimuli are exploited to realize magnetization reversals required for skyrmion formation such as magnetic field~\cite{Koshibae15,Buttner15,Mochizuki17}, electric current~\cite{Iwasaki13b,Sampaio13,ZhouY14,Everschor17,JiangW15,SWoo16,Buttner17,SWoo18}, electric field~\cite{Mochizuki15a,Mochizuki15b,Okamura16,Schott17,MaC19,Huang18,Kruchkov18,WangL18}, laser light~\cite{Finazzi13,Koshibae14,Berruto18,JeSG18}, and many others~\cite{Romming13,Oike16,LiuY15,Nii15}. Basically, it is much easier to create skyrmion lattices and skyrmion bubbles compared with nanometric single skyrmions because it is not necessary to squeeze the field-applied spot finely to a nanometric area for the lattice and bubble forms of skyrmions. Here the skyrmion lattices are crystallized forms of skyrmions, and the skyrmion bubbles are magnetic bubbles characterized by a skyrmion-like magnetization configuration with a large size of micrometer order. However, to exploit magnetic skyrmions as information carriers in memory devices, it is necessary to establish a technique to create single skyrmions at an intended nanometric area with a low cost in energy. However, there has been only a few successful experimental demonstrations of the controlled creation of nanometric single skyrmions in spite of a lot of theoretical proposals. This is because it is technically difficult to squeeze the external stimulus to the size of a nanometer spot. One previous experiment succeeding in the one-by-one creation of nanometric skyrmions is based on the spin-polarized current injection via a tip electrode~\cite{Romming13}, which is, however, not an easy technique because it requires equipment of the scanning tunneling microscopy. Theoretically, a lot of methods have been proposed so far such as methods based on the magnetic-field application~\cite{Mochizuki17} and the electric-current injection~\cite{Iwasaki13b} to a thin-plate sample of chiral-lattice ferromagnet with fabricated nanometer-sized notch(es) or hole(s) and methods based on the local electric-field application with a needle-shaped electrode to a thin-plate sample of chiral-lattice multiferroics~\cite{Mochizuki15a,Mochizuki15b}.

In this paper, we propose from a theoretical perspective a controlled technique to create magnetic skyrmions using irradiation with a circularly polarized microwave magnetic field to a thin-plate sample of chiral-lattice magnet with fabricated two rectangular holes [see Fig.~\ref{Fig01}(b)]. We first discuss how a circularly polarized magnetic field effectively induces a steady magnetic field perpendicular to the plane of the circular microwave polarization, where its magnitude is proportional to the microwave frequency $\Omega=2\pi f$. Next, we perform micromagnetic simulations based on the Landau-Lifshitz-Gilbert (LLG) equation to simulate the magnetization dynamics and the skyrmion creation processes under application of a circularly polarized microwave field. Our method exploits intense interferences of microwave-activated spin waves caused by the rectangular holes to realize local magnetization flop to create skyrmion seeds. Advantages of our method are (1) we can create skyrmions in an intended nanometric area between the fabricated holes without squeezing an applied microwave field and (2) we can create individual skyrmions efficiently with a low cost in energy by exploiting the microwave-active resonant excitations of magnetization. Recently, experimental techniques for circularly polarized microwave irradiation have been developed that strongly support the realization of our proposal~\cite{Arakawa19}.

\section{Theory of the Physical Mechanism}
We discuss how a rotating magnetic field effectively induces a steady magnetic-field component perpendicular to the rotation plane. Here we consider a dimensionless magnetization vector $\bm m \equiv -\bm S/\hbar$ with $\bm S$ being a spin vector. The Zeeman coupling between $\bm m$ and a rotating magnetic field $\bm H_\Omega$ is given by,
\begin{eqnarray}
\mathcal{H}_{\rm Zeeman}=-\gamma\hbar \bm H_\Omega \cdot \bm m,
\label{eq:derivA}
\end{eqnarray}
where $\gamma(=g\mu_{\rm B}/\hbar)$ is the electron gyromagnetic ratio. From this Hamiltonian, we derive the time evolution equation for $\bm m$ with respect to the rest coordinates,
\begin{eqnarray}
\frac{d\bm m}{dt}=-\gamma \bm m \times \bm H_\Omega.
\label{eq:derivB}
\end{eqnarray}

The rotating magnetic field $\bm H_\Omega$ is represented by $\bm H_\Omega=H_\Omega(\cos\Omega t, \sin\Omega t, 0)$ with respect to the rest coordinates, the magnetization vector $\bm m$ written in the coordinates rotating with this field is
\begin{eqnarray}
\bm m=m_\xi \bm e_\xi + m_\eta \bm e_\eta + m_\zeta \bm e_\zeta,
\label{eq:deriv1}
\end{eqnarray}
where $\bm e_\xi$, $\bm e_\eta$ and $\bm e_\zeta$ are unit directional vectors of the rotating coordinate system [see Fig.~\ref{Fig01}(c)], which are represented with respect to the rest coordinates as $\bm e_\xi=(\cos\Omega t, \sin\Omega t, 0)$, $\bm e_\eta=(-\sin\Omega t, \cos\Omega t, 0)$, and $\bm e_\zeta=(0, 0, 1)$. The $\xi$-axis is chosen parallel to the $\bm H_\Omega$ field. Taking the time-derivative of both sides of Eq.~(\ref{eq:deriv1}), we obtain
\begin{eqnarray}
\left. 
\frac{d\bm m}{dt}=\frac{d\bm m}{dt} \right|_{\rm R}
+\Omega \left(m_\xi \bm e_\eta - m_\eta \bm e_\xi \right),
\label{eq:deriv2}
\end{eqnarray}
where $(\ldots)_{\rm R}$ denotes a vector expressed in the rotating coordinates with bases $\bm e_\xi$, $\bm e_\eta$ and $\bm e_\zeta$. In the derivation, we used relations
\begin{eqnarray}
& &
\left. 
\frac{d\bm m}{dt} \right|_{\rm R}
=\frac{dm_\xi}{dt} \bm e_\xi + \frac{dm_\eta}{dt} \bm e_\eta 
+ \frac{dm_\zeta}{dt} \bm e_\zeta, 
\\
& &
\frac{d\bm e_\xi}{dt}=\Omega \bm e_\eta, 
\quad
\frac{d\bm e_\eta}{dt}=-\Omega \bm e_\xi,
\quad
\frac{d\bm e_\zeta}{dt}=0.
\label{eq:deriv3}
\end{eqnarray}
Substituting the following two relations,
\begin{eqnarray}
\frac{d\bm m}{dt}=-\gamma \bm m \times \bm H_\Omega=-\gamma
\begin{pmatrix}m_\xi \\ m_\eta \\ m_\zeta\end{pmatrix}_{\rm R}
\times
\begin{pmatrix}H_\Omega \\ 0 \\ 0\end{pmatrix}_{\rm R}
\label{eq:deriv4}
\end{eqnarray}
and
\begin{eqnarray}
\Omega \left(m_\xi \bm e_\eta - m_\eta \bm e_\xi \right)=
\begin{pmatrix}-\Omega m_\eta \\ \Omega m_\xi \\ 0\end{pmatrix}_{\rm R}
=
\begin{pmatrix}m_\xi \\ m_\eta \\ m_\zeta\end{pmatrix}_{\rm R}
\times
\begin{pmatrix}0 \\ 0 \\ -\Omega\end{pmatrix}_{\rm R}
\label{eq:deriv5}
\end{eqnarray}
into Eq.~(\ref{eq:deriv3}), we obtain
\begin{eqnarray}
\begin{pmatrix}dm_\xi/dt \\ dm_\eta/dt \\ dm_\zeta/dt \end{pmatrix}_{\rm R}
=-\gamma
\begin{pmatrix}m_\xi \\ m_\eta \\ m_\zeta\end{pmatrix}_{\rm R}
\times
\begin{pmatrix}H_\Omega \\ 0 \\ -\Omega/\gamma\end{pmatrix}_{\rm R}.
\label{eq:deriv6}
\end{eqnarray}
This equation indicates that the circularly polarized in-plane magnetic field effectively acts as a magnetic field given by $\bm H_{\rm R}=(H_\Omega, 0, -\Omega/\gamma)$, which contains a steady out-of-plane component proportional to angular frequency $\Omega$ [see Fig.~\ref{Fig01}(d)] and therefore induces a magnetization component perpendicular to the plane. In this way, a rotating microwave magnetic field $\bm H_\Omega$ induces or switches the magnetization. The physical origin of the perpendicular steady field component is an angular momentum transfer from the rotating magnetic field to the magnetic moment, and the above argument based on the coordinate translation successfully provides its formulation. In fact, this kind of argument is often used to explain the principles of magnetic-moment switching by $\pi$-pulse or $\pi/2$-pulse in the nuclear magnetic resonance (NMR) measurements~\cite{NMR1,NMR2}.

\section{Details of Numerical Simulations}
To demonstrate the skyrmion creation with a circularly polarized microwave magnetic field, we perform micromagnetic simulations using the LLG equation,
\begin{equation}
\frac{d\bm m_i}{dt}=-\frac{\gamma}{1+\alpha^2}
\left[
\bm m_i \times \bm H^{\rm eff}_i 
+\frac{\alpha}{m} \bm m_i \times \left(\bm m_i \times 
\bm H^{\rm eff}_i \right)
\right].
\label{eq:LLGEQ}
\end{equation} 
Here $\bm m_i$($\equiv -\bm S_i\hbar$) represents a dimensionless classical magnetization vector on the $i$th site of the square lattice where $\bm S_i$ is the spin vector. The norm $m=|\bm m_i|$ is fixed at unity. The first term of the right-hand side describes the precession of local magnetization on the $i$th site around the effective magnetic field $\bm H_i^{\rm eff}$, which can be calculated from the Hamiltonian $\mathcal{H}$ as
\begin{equation}
\bm H^{\rm eff}_i = -\frac{1}{\gamma\hbar}\frac{\partial \mathcal{H}}{\partial \bm m_i}.
\label{eq:EFFMF}
\end{equation}
The second term is a phenomenologically introduced damping term in which $\alpha$(=0.04) is the Gilbert-damping coefficient. 

The Hamiltonian is composed of two terms, $\mathcal{H}$=$\mathcal{H}_0$+$\mathcal{H}^\prime(t)$. The first term $\mathcal{H}_0$ is the model Hamiltonian for a thin-plate specimen of a chiral-lattice magnet. This term contains the ferromagnetic-exchange interaction, the DM interaction, and the Zeeman interaction associated with a static external magnetic field $\bm H_{\rm ex}=(0,0,H_{\rm ex})$ applied perpendicular to the sample plane. This Hamiltonian is given by~\cite{Bak80,YiSD09},
\begin{eqnarray}
\mathcal{H}_0
&=&-J \sum_{i} (\bm m_i \cdot \bm m_{i+\hat{x}}+\bm m_i \cdot \bm m_{i+\hat{y}})\nonumber \\
& &-D \sum_{i} (\bm m_i \times \bm m_{i+\hat{x}} \cdot \hat{\bm x}
+\bm m_i \times \bm m_{i+\hat{y}} \cdot \hat{\bm y})
\nonumber \\
& &-\gamma\hbar H_{\rm ex} \sum_i m_{zi}.
\label{eq:model}
\end{eqnarray}
The strength of the DM interaction is fixed at $D/J$=0.27, whereas the intensity of the static external magnetic field is fixed at $\gamma\hbar H_{\rm ex}/J$=0.057.

Note that we neglect the magnetic dipole interaction in the present study because we know that it does not affect both static and dynamical behaviors of the nanometric skyrmions for the following two grounds. First, we can theoretically argue that the sample thickness must be unphysically thin, i.e., less than 1 nm, in order to create skyrmions with a diameter of approximately 10 nm by the magnetic dipole interaction~\cite{Iwasaki13a}. This indicates that for samples thicker than 1 nm, which is usually the case for real experiments, we can neglect the effects of magnetic dipole interaction on the static properties of nanometric skyrmions. Second, a recent work based on the combined experimental and theoretical studies in Ref.~\cite{Schwarze15} confirmed that the resonance frequencies, the eigenmodes and even the excitation amplitudes of microwave-activated magnetic skyrmions in a chiral-lattice magnet Cu$_2$OSeO$_3$ change only little upon the sample-shape variation among thin-plate, stick-shaped, and three-dimensional bulk samples, indicating that the magnetostatic energy due to the magnetic dipole interaction does not affect the resonant dynamics of nanometric skyrmions. Thus, we employ a simple Hamiltonian without magnetic dipole interaction as a minimal model in this study.

The second term $\mathcal{H}'(t)$ describes the coupling between magnetizations $\bm m_i$ and a circularly polarized microwave magnetic field $\bm H_\Omega=H_\Omega(\cos\Omega t, \sin\Omega t, 0)$,
\begin{equation}
\mathcal{H}^\prime(t)=-\gamma\hbar\bm H_\Omega(t) \cdot \sum_i \bm m_i.
\label{eq:TDEPH}
\end{equation}
Here we introduce dimensionless quantities, $\tau$, $\omega$ and $h$, respectively for time, angular frequency, and magnetic field as,
\begin{equation}
\tau=\frac{tJ}{\hbar},\hspace{0.5cm} 
\omega=\hbar\Omega/J,\hspace{0.5cm} 
h=\gamma\hbar H/J.
\end{equation}
Using these dimensionless quantities, we rewrite the term $\mathcal{H}^\prime$ as
\begin{equation}
\mathcal{H}^\prime(\tau)=-\bm h_\omega(\tau)J \cdot \sum_i \bm m_i,
\label{eq:TDEPH2}
\end{equation}
with $\bm h_\omega=h_\omega(\cos\omega\tau, \sin\omega\tau, 0)$ and $h_\omega=\gamma\hbar H_\Omega/J$.

We analyze the LLG equation in Eq.~(\ref{eq:LLGEQ}) using the fourth-order Runge-Kutta method and trace the spatiotemporal dynamics of the magnetizations. The unit conversions when $J$=1~meV are summarized in Table~\ref{tab:uconv}.
\begin{table}
\caption{Unit conversion table when $J$=1~meV.}
\begin{tabular}{l|cc} \hline \hline
               & Dimensionless & Corresponding value \\
               & quantity & with units \\
\hline
Exchange int.  & \hspace{0.5cm} $J$=1        & \hspace{0.5cm} $J$=1~meV \\
Time           & \hspace{0.5cm} $\tau \equiv tJ/\hbar$=1 & \hspace{0.5cm} $t$=0.66~ps \\
Frequency $f=\Omega/2\pi$ &\hspace{0.5cm} 
$\omega \equiv \hbar\Omega/J$=1 & \hspace{0.5cm} $f=\Omega/2\pi$=241~GHz \\
Magnetic field & \hspace{0.5cm} $h \equiv \gamma\hbar H/J$=1 & 
\hspace{0.5cm} $H$=8.64~T \\ 
\hline \hline
\end{tabular}
\label{tab:uconv}
\end{table}

\section{Results}
\begin{figure}[t]
\includegraphics[width=1.0\columnwidth]{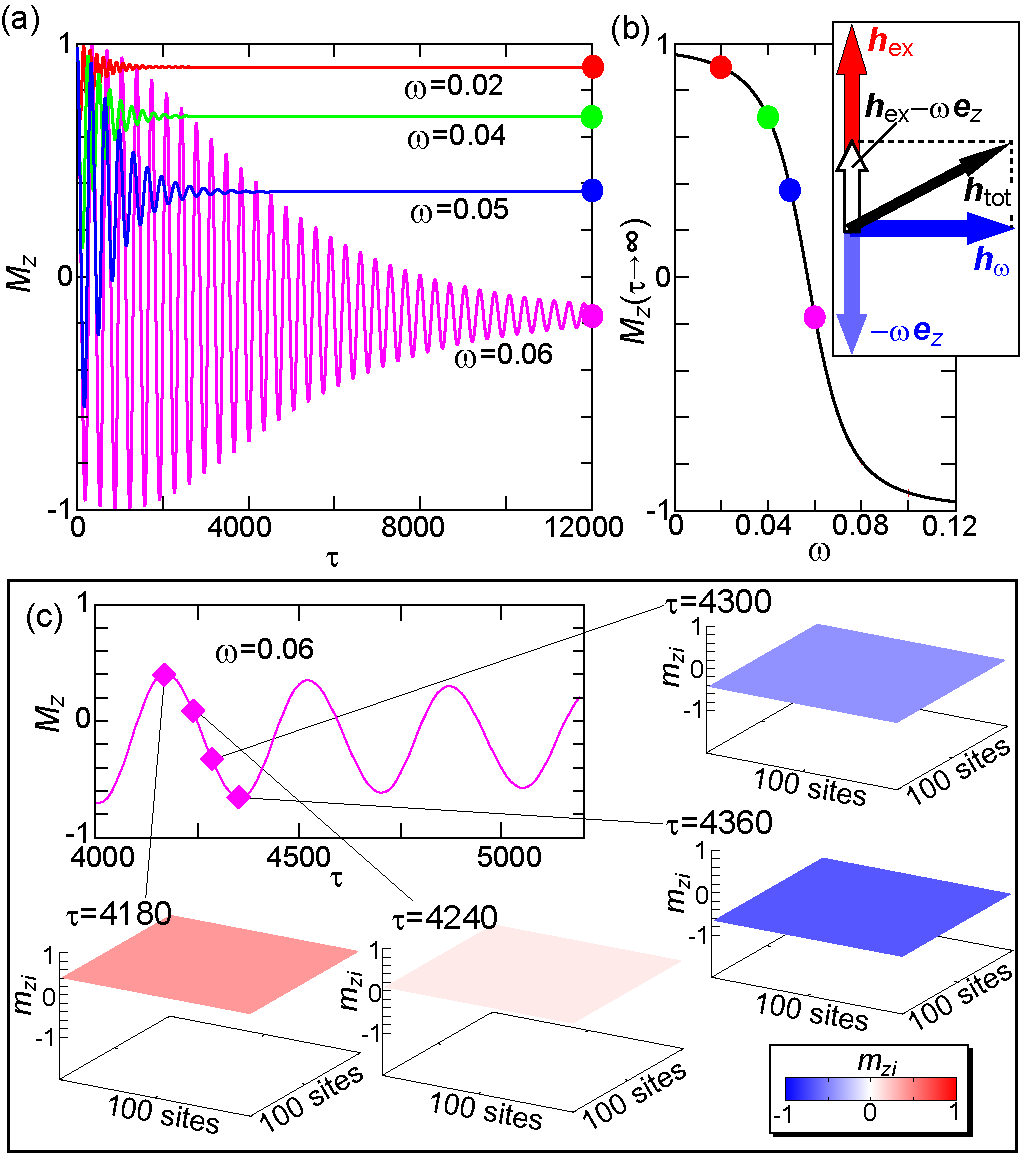}
\caption{(a) Simulated time profiles of the net magnetization $M_z$ under application of a circularly polarized microwave field $\bm h_\omega=h_\omega(\cos\omega \tau, \sin\omega \tau, 0)$ with $h_\omega$=0.018 to a field-polarized ferromagnetic state for various microwave frequencies $\omega$. (b) Microwave-frequency dependence of the saturated net magnetization $M_z(\tau\rightarrow\infty)$ after a sufficient duration of microwave irradiation. The solid line indicates the theoretical prediction of the behavior of $M_z^{\rm sat}$ given by Eq.~(\ref{eq:SatMag}). The inset shows the effective magnetic field $\bm h_{\rm tot}$ acting on the magnetizations, the in-plane component of which comes from the rotating microwave magnetic field, whereas its out-of-plane component is composed of two contributions: the static external field $\bm h_{\rm ex}$ and the effective steady field $-\omega \bm e_z$ induced by the circularly polarized microwave. (c) Snapshots of the spatial profiles of temporally oscillating $m_{zi}$ at selected moments when $\omega$=0.06.}
\label{Fig02}
\end{figure}
We first simulate magnetization dynamics when a circularly polarized microwave field $\bm h_\omega$ with $h_\omega$=0.018 is applied to a uniformly field-polarized ferromagnetic state in a system without any spatial structures such as notches and holes. Here we use a system of $N=100 \times 100$ sites with periodic boundary conditions. Figure~\ref{Fig01}(a) shows the simulated time profiles of the out-of-plane component $M_z$ of net magnetization $\bm M$=$(1/N)\sum_i \bm m_i$ for various microwave frequencies $\omega$. The continuous irradiation with the microwave is started at $\tau$=0. We find that the magnetization $M_z$ shows a damped oscillation that saturates to a certain finite value. We plot in Fig.~\ref{Fig02}(b) the saturation values $M_z^{\rm sat}$ after sufficient duration as a function of the angular frequency $\omega$.

In the present instances, the magnetic field $\bm h_{\rm tot}$ acting on local magnetizations is composed of two contributions: the static external magnetic field $\bm h_{\rm ex}$ and the rotating microwave magnetic field $\bm h_\omega$. More specifically, the in-plane component of $\bm h_{\rm tot}$ corresponds to the in-plane component of $\bm h_\omega$, whereas the out-of-plane component of $\bm h_{\rm tot}$ is a sum of the external magnetic field $\bm h_{\rm ex}$ and an effective steady component $-\omega\bm e_z$ induced by the rotating microwave field [see inset of Fig.~\ref{Fig02}(b)]. Consequently, the total magnetic field $\bm h_{\rm tot}$ is given by,
\begin{equation}
\bm h_{\rm tot}=(h_\omega \cos\omega \tau, \;h_\omega \sin\omega \tau, \;
h_{\rm ex}-\omega),
\label{eq:Htot}
\end{equation}
Note that the out-of-plane component $h_{\rm ex}-\omega$ is time-independent, whereas the in-plane components $h_\omega(\cos\omega\tau, \sin\omega\tau)$ are dynamical. The net magnetization $\bm M$ temporally changes its direction to follow this dynamical magnetic field $\bm h_{\rm tot}$, but its out-of-plane component becomes steady when the irradiated system assumes a nonequilibrium steady state after sufficient duration.

The saturated out-of-plane component $M_z^{\rm sat}$ in the nonequilibrium steady state is given by 
\begin{equation}
M_z^{\rm sat}=\frac{h_{\rm ex}-\omega}
{\sqrt{(h_{\rm ex}-\omega)^2+h_\omega^2}}.
\label{eq:SatMag}
\end{equation}
The solid line in Fig.~\ref{Fig02}(b) depicts the behavior of this magnetization component and perfectly coincides with the values of $M_z^{\rm sat}$ from simulations. This indicates that the local magnetizations $\bm m_i$ temporally oscillate uniformly in space. Indeed, simulated snapshots of the spatial profiles of the temporally oscillating $m_{zi}$ at selected moments in Fig.~\ref{Fig02}(c) show spatially uniform oscillations of $\bm m_i$ in the transient stage before magnetization saturation. To create magnetic skyrmions, however, it is necessary to reverse the magnetizations locally to form skyrmion cores. This means that applying a circularly polarized microwave field cannot create skyrmions as long as spatially uniform systems are considered, as in the present simulations.

\begin{figure}[t]
\includegraphics[width=1.0\columnwidth]{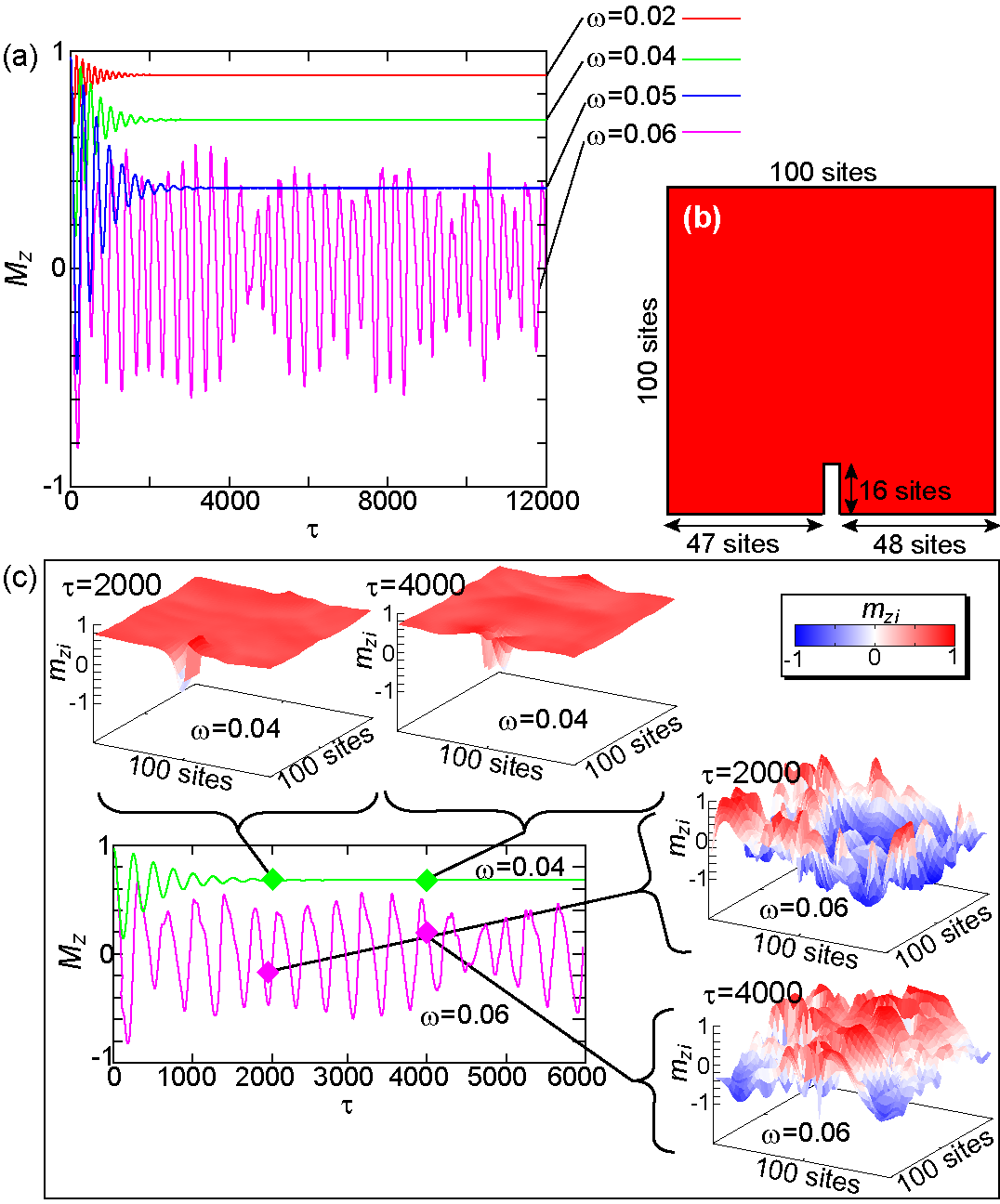}
\caption{(a) Simulated time profiles of the net magnetization $M_z$ under application of a circularly polarized microwave field with $h_\omega$=0.018 to a ferromagnetic system with a rectangular notch for various microwave frequencies $\omega$. (b) Field-polarized ferromagnetic sample with a rectangular notch the size of which encompasses 5 $\times$ 16 sites used in the micromagnetic simulations, on which the periodic boundary conditions are imposed. (c) Snapshots of the spatial profiles of temporally oscillating $m_{zi}$ at selected times for $\omega$=0.04 and $\omega$=0.06.}
\label{Fig03}
\end{figure}
We next simulate magnetization dynamics when a circularly polarized microwave field $\bm h_\omega(\tau)$ with $h_\omega$=0.018 is applied to a ferromagnetic system with a fabricated structure. Figure~\ref{Fig03}(a) shows the simulated time profiles of $M_z$ for various microwave frequencies $\omega$ when a system has a rectangular notch encompassing 5 $\times$ 16 sites [see Fig.~\ref{Fig03}(b)]. When the microwave frequency $\omega$ is small as $\omega \leq 0.05$, they behave as damped oscillations saturating to a certain value similar to those in Fig.~\ref{Fig02}(a). In contrast, the behaviors are entirely different when $\omega$ is large as $\omega=0.06$. They show beating and even chaos, which can be attributed to complicated reflections and interferences of excited spin waves caused by the notch. In Fig.~\ref{Fig03}(c), simulated snapshots of the spatial profiles of temporally oscillating $m_{zi}$ at selected moments are shown for $\omega$=0.04 and $\omega$=0.06. We find that in contrast to a rather spatially uniform distribution of $m_{zi}$ for a small $\omega$(=0.04), it shows a complicated distribution for a large $\omega$(=0.06). More importantly, both negative and positive $m_{zi}$ coexist in the latter indicating that local reversals of magnetizations take place when the amplitudes of magnetization oscillations are locally amplified by a significant interference effect of the spin waves. Note that this effect is caused purely by the introduced notch structure but not by the system edge because the periodic boundary conditions are again imposed and thus edges are absent in this system.

\begin{figure}[t]
\includegraphics[width=1.0\columnwidth]{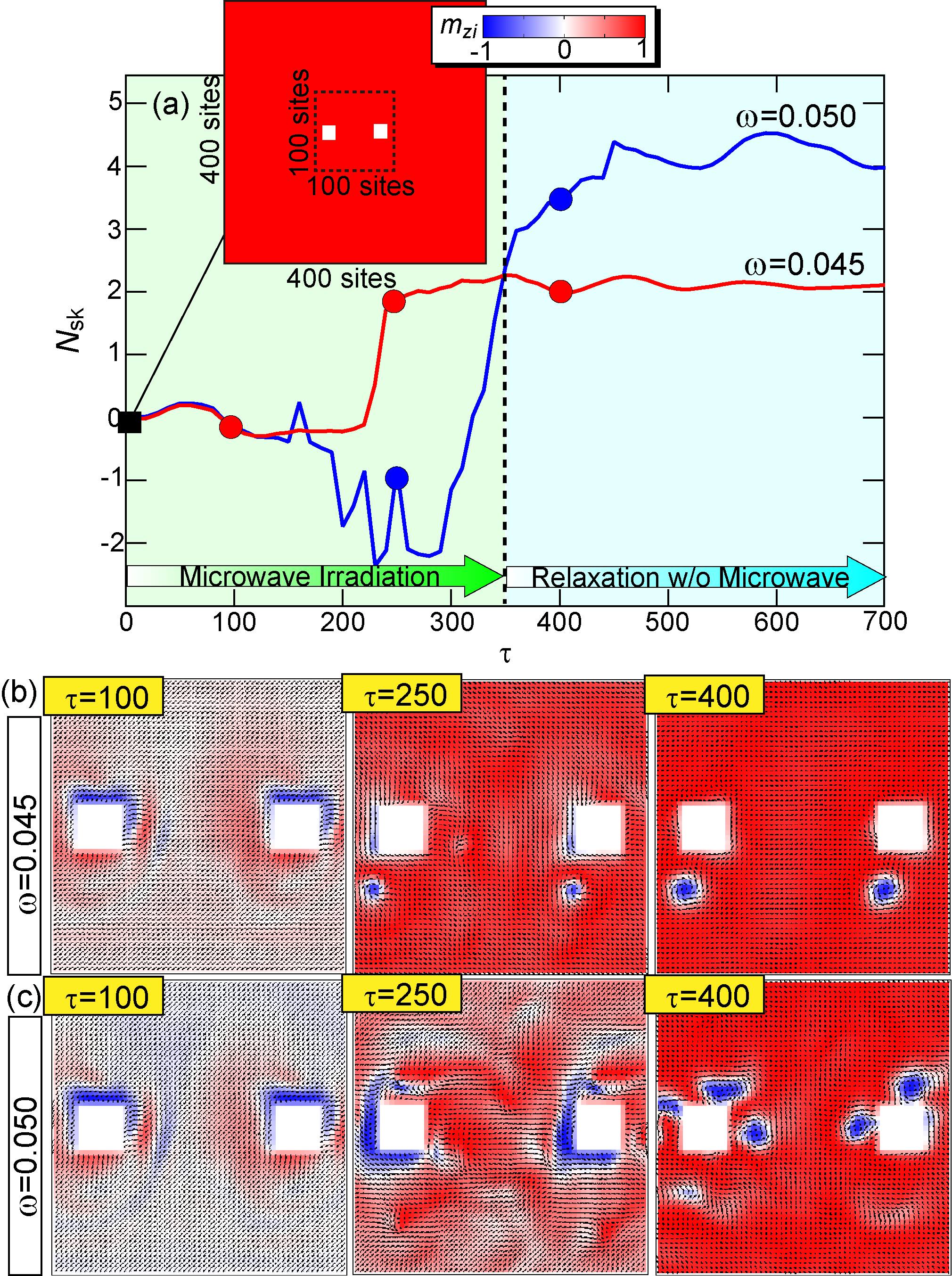}
\caption{(a) Simulated time profiles of the skyrmion number $N_{\rm sk}$ when a circularly polarized microwave field $\bm h_\omega=h_\omega(\cos\omega\tau, \sin\omega\tau, 0)$ with $h_\omega$=0.018 is applied to a field-polarized ferromagnetic system with two rectangular holes for $\omega$=0.045 and $\omega$=0.05. A system of 400$\times$400 sites with rectangular holes (see inset) is used for the simulations. Microwave irradiation starts at $\tau$=0 and ends at $\tau$=300. (b), (c) Snapshots of the skyrmion-creation processes for (b) $\omega$=0.045 and (c) $\omega$=0.05. Here we focus on an area encompassing 100$\times$100 sites around the holes indicated by a dashed square in the inset of (a).}
\label{Fig04}
\end{figure}
The local magnetization reversals caused by the spin-wave interferences indeed result in the creation of magnetic skyrmions. We simulated the creation processes by application of a circularly polarized microwave field $\bm h_\omega(\tau)$ with $h_\omega$=0.018 to a field-polarized ferromagnetic state in a system with two rectangular holes [see inset of Fig.~\ref{Fig04}(a)]. The skyrmion formation is detectable by calculating the skyrmion number $N_{\rm sk}$. The skyrmions are characterized by a quantized topological invariant which corresponds to a sum of solid angles spanned by three neighboring magnetization vectors. One skyrmion has a topological invariant of $-4\pi$ because the magnetization vectors constituting a skyrmion point in every direction wrapping a sphere once. Therefore, the quantity $N_{\rm sk}$ is calculated by,
\begin{equation}
N_{\rm sk}=-\frac{1}{4\pi}\sum_i
(\bm m_{i+\hat{\bm x}} \times \bm m_{i+\hat{\bm y}})\cdot \bm m_i.
\label{eq:Nsk}
\end{equation}

Figure~\ref{Fig04}(a) shows the simulated time profiles of $N_{\rm sk}$ for different microwave frequencies, $\omega$=0.045 and $\omega$=0.05. Here the system is irradiated with a continuous microwave field from $\tau$=0 to $\tau$=300, after which the irradiation is stopped. The skyrmion number $N_{\rm sk}$ starts increasing immediately afterwards, indicating that skyrmions are created, and after a sufficient duration converges to a finite value. In Figs.~\ref{Fig04}(b) and (c), we display simulated snapshots of the magnetization distributions at selected moments during the skyrmion creation for two different microwave frequencies, specifically, (b) $\omega$=0.045 and (c) $\omega$=0.05. The applied microwave field activates spin waves of the ferromagnetically aligned magnetizations, and the excited spin waves propagate and are reflected by the fabricated holes. The reflected spin waves mutually interfere in complicated manners that result in spatially inhomogeneous amplitudes of the magnetization oscillation. In consequence, both areas with positively oriented magnetizations and those with negatively oriented magnetizations emerge and coexist as seen in the snapshots at $\tau$=250. After the microwave irradiation is stopped, the system gradually relaxes and most of the magnetizations become reoriented along the external magnetic field $\bm h_{\rm ext}$. However, some of the areas with reversed magnetizations remain to seed skyrmions. After sufficient relaxation, these skyrmion seeds grow to be skyrmion cores, resulting in the skyrmion formation.

It should be mentioned that the area sandwiched by the holes is an important parameter rather than the size of holes for successful skyrmion creation in this technique. We need to prepare an area larger than the skyrmion diameter, which is determined by the ratio $D/J$ where $D$ and $J$ are strengths of the DM interaction and the ferromagnetic exchange interaction, respectively. This is because this area should be large enough to host skyrmion seeds with locally reversed magnetizations. But, at the same time, the area should be small enough to realize efficient interferences of spin waves before they are damped. Therefore, we use a system in which the distance between two faced edges of the rectangular holes is set to be 60 sites. This distance is twice larger than the skyrmion diameter of 33 sites with the ratio $D/J$=0.27 used in the present study. 

\begin{figure}[t]
\includegraphics[width=1.0\columnwidth]{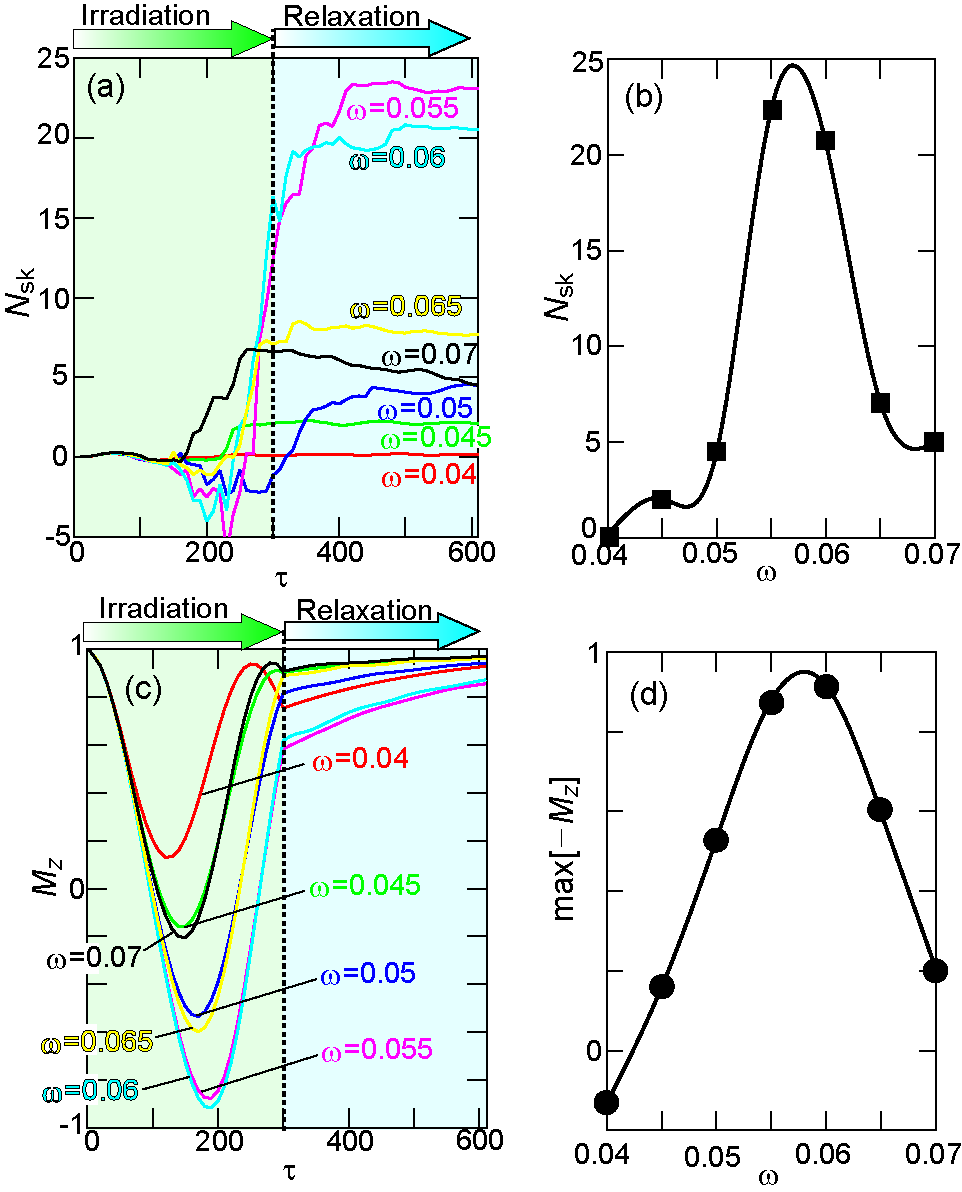}
\caption{(a),(c) Simulated time profiles of (a) the skyrmion number $N_{\rm sk}$ and (c) the net magnetization $M_z$ when a circularly polarized microwave field $\bm h_\omega=h_\omega(\cos\omega\tau, \sin\omega\tau, 0)$ with $h_\omega$=0.018 is applied to a field-polarized ferromagnetic system with two rectangular holes [see inset of Fig.~\ref{Fig04}(a)] for various microwave frequencies $\omega$. Microwave irradiation starts at $\tau$=0 and ends at $\tau$=300. (b) Microwave-frequency dependence of $N_{\rm sk}$ at $\tau$=600 after sufficient relaxation. (d) Microwave-frequency dependence of maximum value of $-M_z(\tau)$.}
\label{Fig05}
\end{figure}
Figure~\ref{Fig05}(a) shows simulated time profiles of $N_{\rm sk}$ when a circularly polarized microwave field with $h_\omega$=0.018 is applied to the system [inset of Fig.~\ref{Fig04}(a)] for various microwave frequencies $\omega$. Again, we trace the magnetization dynamics for both the activation process under application of the microwave field from $\tau$=0 to $\tau$=300 and the subsequent relaxation with no microwave field from $\tau$=300. We find that the number of created skyrmions $N_{\rm sk}$ after a sufficient relaxation is not monotonically dependent on $\omega$, although the steady perpendicular component of the effective magnetic field $h_{\rm ex}-\omega$ is linearly dependent on $\omega$. This non-monotonic $\omega$ dependence is clearly seen in the plot of $N_{\rm sk}$ at $\tau$=600 as a function of $\omega$ in Fig.~\ref{Fig05}(b), which indicates that $N_{\rm sk}$ has a maximum at $\omega\sim0.055$. 

To understand this non-monotonic behavior and maximum peak, we investigated time profiles of $M_z$ by varying the microwave frequency $\omega$ [see Fig.~\ref{Fig05}(c)]. We find that the value of $M_z$ starts decreasing right after the microwave irradiation sets in because the microwave field effectively reduces the steady out-of-plane component of the magnetic field from $h_{\rm ex}$ to $h_{\rm ex}-\omega$. The magnetization $M_z$ has a minimum at a certain instant and then increases. This behavior covers the first half-cycle of the damped oscillation of $M_z$. Note that if we continue the irradiation, the magnetization $M_z$ should saturate to a certain value after the damped oscillations or should keep oscillating [Fig.~\ref{Fig03}(a)]. However, the irradiation is stopped before completion of the first cycle of oscillation. In this initial oscillation process, its amplitude is governed by the sensitivity or susceptibility of the system to the microwave magnetic field, and it tends to be enhanced when the microwave has a frequency close to the ferromagnetic resonance frequency. The previous theoretical study revealed that the ferromagnetic resonance is located around $\omega\sim0.055$ when the strength of the DM interaction is $D/J$=0.27 as used in the present study~\cite{Mochizuki12}. In Fig.~\ref{Fig05}(d), we plotted the maximum value of $-M_z$ in the initial oscillation process under microwave irradiation, which represents the extent of the microwave-induced magnetization reversal. We see a peak around $\omega$=0.055, indicating that the applied microwave field reverses the magnetizations efficiently when $\omega$ is close to the ferromagnetic resonance. This efficient magnetization reversal leads to an abundant creation of skyrmions and hence the observed peak in $N_{\rm sk}$ around $\omega=0.055$ [Fig.~\ref{Fig05}(b)].

Note that the magnetization dynamics during the skyrmion-creation process is in a strongly nonlinear regime because of the large amplitude of magnetization oscillations induced by the strong microwave magnetic field. However, we find that the microwave frequency for the most efficient skyrmion creation coincides with the ferromagnetic resonance frequency in the linear regime predicted in the previous theoretical study~\cite{Mochizuki12}. A possible reason is as follows. When the magnetizations start oscillating in the very beginning process of the magnetization switching, their dynamics is in the linear regime because of the small amplitude of oscillation. We expect that the efficiency of skyrmion creation is governed by whether the applied microwave field can induce this initial magnetization oscillation efficiently because once the magnetizations could start oscillating, the oscillation amplitude can be easily enhanced by their inertial behaviors. Hence, the most efficient skyrmion creation occurs when the microwave frequency is tuned at the ferromagnetic resonance frequency in the linear regime.

\begin{figure*}
\includegraphics[width=2.0\columnwidth]{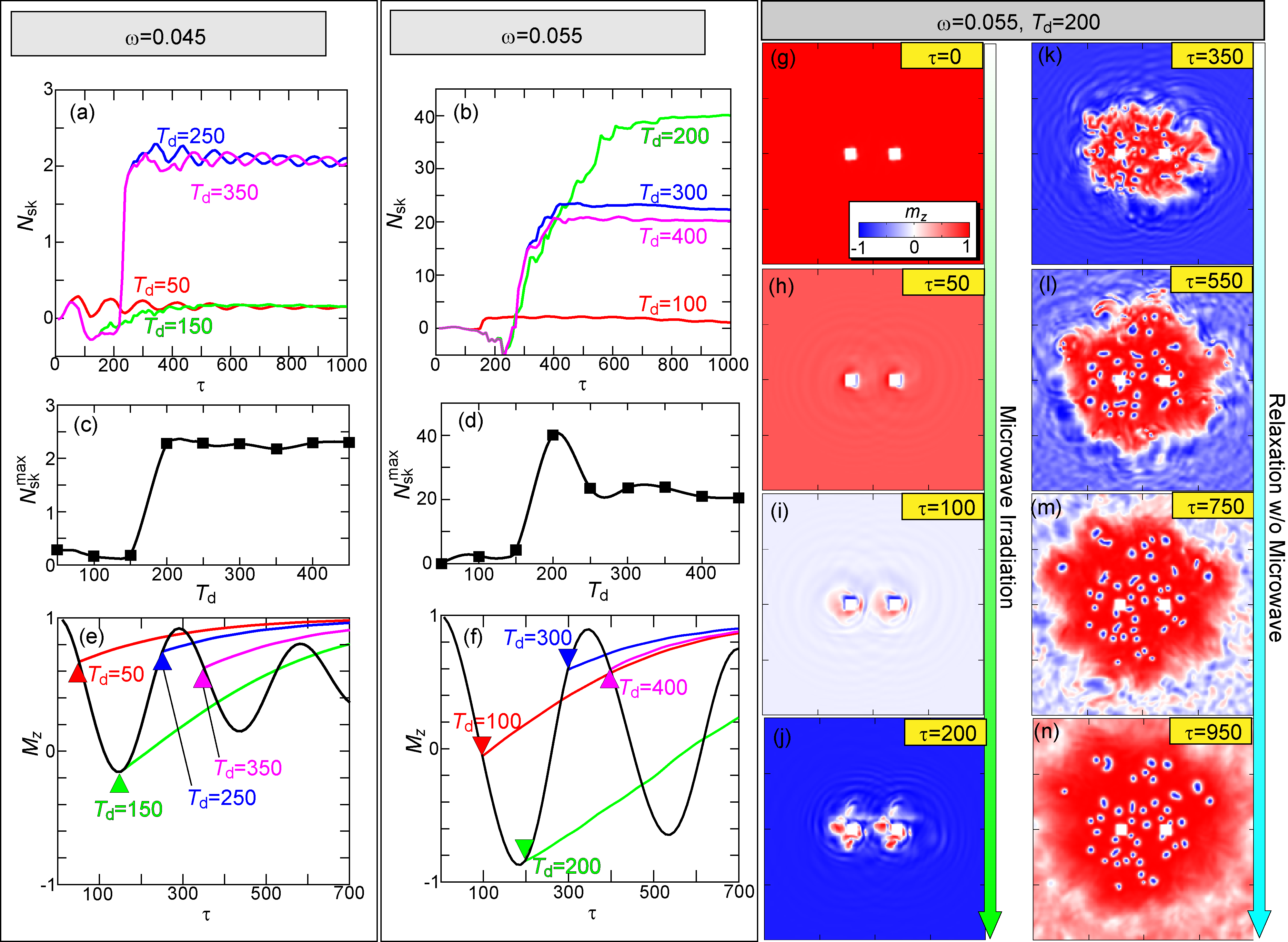}
\caption{(a), (b) Simulated time profiles of the skyrmion number $N_{\rm sk}$ for different microwave frequencies, i.e., (a) $\omega$=0.045 and (b) $\omega$=0.055. (c), (d) Dependence of the maximum value of the skyrmion number $N_{\rm sk}^{\rm max}$ on the microwave duration $T_{\rm d}$ for (c) $\omega$=0.045 and (d) $\omega$=0.055. (e), (f) Simulated time profiles of the net magnetization $M_z$ for (e) $\omega$=0.045 and (f) $\omega$=0.055. (g)--(n) Simulated snapshots, in which numerous skyrmions are created, of the spatiotemporal dynamics of their magnetizations under application of a circularly polarized microwave field $\bm h_\omega=h_\omega(\cos\omega\tau, \sin\omega\tau, 0)$ with $h_\omega=0.018$ when $\omega$=0.055 and $T_{\rm d}$=200. Here (g)--(j) correspond to activation period under the application of a microwave field for $0 \leq \tau \leq T_{\rm d}$, whereas (k)--(n) correspond to subsequent relaxation without microwave irradiation for $\tau>T_{\rm d}$.}
\label{Fig06}
\end{figure*}
We also studied the microwave-duration dependence of skyrmion creation. The magnetization dynamics were simulated for the activation period during which a microwave field $\bm h_\omega$ with $h_\omega=0.018$ is applied from $\tau$=0 to $\tau$=$T_{\rm d}$ and the subsequent relaxation with no irradiation after $\tau$=$T_{\rm d}$ for various lengths of duration $T_{\rm d}$. Two instances with different microwave frequencies, specifically, $\omega$=0.045 and $\omega$=0.055, were examined. The simulated time profiles of $N_{\rm sk}$ are shown in Figs.~\ref{Fig06}(a) and (b) for $\omega$=0.045 and $\omega$=0.055, respectively, whereas in Figs.~\ref{Fig06}(c) and (d), the maximum values of the skyrmion number $N_{\rm sk}^{\rm max}$ are plotted as functions of $T_{\rm d}$ for the respective frequencies. For a small microwave frequency of $\omega$=0.045, the observed $N_{\rm sk}^{\rm max}$ is almost zero when the duration is rather short ($T_{\rm d} \leq 150$), whereas it nearly saturates to a certain constant value of $\sim 2$ when the duration is rather long ($T_{\rm d} \geq 200$) [see Fig.~\ref{Fig06}(c)]. In contrast, when $\omega$ is rather large ($\omega$=0.055), the behavior of $N_{\rm sk}^{\rm max}$ is not monotonic but features a peak around a certain value of $T_{\rm d}$ [Fig.~\ref{Fig06}(d)]. Surprisingly [see Fig.~\ref{Fig06}(b)], the number of created skyrmion increases explosively as time passes when $T_{\rm d}$=200. 

We now briefly summarize the observations obtained from the above-argued simulation results: (1) the $T_{\rm d}$-dependence of $N_{\rm sk}$ differs between instances with higher and lower $\omega$ as seen when comparing Figs.~\ref{Fig06}(a) and (b); (2) the $T_{\rm d}$-dependence of $N_{\rm sk}^{\rm max}$ is non-monotonic when $\omega$ is large as seen in Fig.~\ref{Fig06}(d); and (3) the number of skyrmions $N_{\rm sk}$ becomes extremely large for a certain value of $T_{\rm d}$ when $\omega$ is large. These observations are understood by considering time profiles of the net magnetization $M_z$ for $\omega$=0.045 [Fig.~\ref{Fig06}(e)] and for $\omega$=0.055 [Fig.~\ref{Fig06}(f)]. The damped sinusoidal oscillations marked by (black) solid lines correspond to the time evolutions of $M_z$ under continuous application of a microwave with nearly infinite or sufficiently long $T_{\rm d}$. In contrast, when $T_{\rm d}$ is short, the magnetization $M_z$ stops its oscillation at $t=T_{\rm d}$ and then gradually returns to its initial fully polarized ferromagnetic state monotonically. Because the microwave-induced effective steady magnetic field $-\omega\bm e_z$ is weak when $\omega$ is small, the value of $M_z$ almost remains positive, and therefore the system is far from displaying a total reversal of magnetizations. Then the skyrmion number $N_{\rm sk}$ scales more or less with duration $T_{\rm d}$. However, when $\omega$ is large, the microwave-induced effective steady magnetic field $-\omega\bm e_z$ is strong and therefore induces an intense oscillation of $M_z$ with a large amplitude [Fig.~\ref{Fig06}(f), (black) solid line]. In this instance, the oscillating $M_z$ takes a large negative value. Indeed, the minimum value of $M_z$ reaches almost $-1$ when $T_{\rm d}$=200, which corresponds to a nearly perfect reversal of magnetizations. If we stop the microwave irradiation when $M_z \sim -1$ at this moment, the system assumes a negatively polarized ferromagnetic state, which reverts back to its initial positively polarized ferromagnetic state. In this transient period, a huge number of skyrmions are created.

Figures.~\ref{Fig06}(g)--(n) display snapshots of the simulated spatiotemporal dynamics of magnetizations under the application of a circularly polarized microwave magnetic field with $\omega$=0.055 and $T_{\rm d}$=200, for which we observe the explosive creation of skyrmions. Here panels (g)--(j) correspond to the activation process under the application of the microwave field during $0 \leq \tau \leq T_{\rm d}$, whereas panels (k)--(n) correspond to subsequent relaxation process without the microwave field when $\tau>T_{\rm d}$. Starting from a positively fully field-polarized ferromagnetic state with $m_{iz}>0$ at $\tau$=0 [Fig.~\ref{Fig06}(g)], the applied circularly polarized microwave field reverses magnetizations over almost the whole area and a negatively polarized ferromagnetic state emerges at $\tau$=200 [Fig.~\ref{Fig06}(j)].

Immediately after the microwave irradiation is stopped at $\tau$=$T_{\rm d}$=200, reorientations of the magnetizations commence the recovery of the initial positively field-polarized ferromagnetic state. The reorientations first occur around the rectangular holes because the magnetizations easily change their orientations at edges of the rectangular holes via gradual and continuous rotations from downwards to upwards. As seen in Fig.~\ref{Fig06}(k)-(n), areas of reoriented magnetizations pointing upwards (red areas) grow and spread quickly. In this transient period, a lot of skyrmion seeds are created. Eventually, we obtain a huge number of skyrmions and skyrmion seeds in the field-polarized ferromagnetic state at $\tau$=950 [Fig.~\ref{Fig06}(n)]. Note that the skyrmion sizes in Figs.~\ref{Fig06}(m) and (n) seem to be distributed. This is because these two figures show transient periods of the skyrmion creation before relaxation. The finally created skyrmions after sufficient duration of relaxation should have the same size of 33 sites determined by the ratio of $D/J$=0.27, which corresponds to 16.5 nm if we assume a typical lattice constant of 5 \AA.

\section{Summary and Discussion}
To summarize, from a theoretical perspective, we have studied a technique to create magnetic skyrmions by applying a circularly polarized microwave field to a ferromagnetically ordered system with DM interactions. We discussed how the circulating microwave magnetic field effectively induces a steady magnetic field component perpendicular to the circular-polarization plane where its magnitude is proportional to the angular frequency $\omega$. We performed micromagnetic simulations based on the LLG equation and demonstrated that nanometric magnetic skyrmions can be created by the application of a microwave to a thin-plate specimen with fabricated rectangular holes. These holes locally amplify the spin waves excited by this effective magnetic field, which induces local reversals of the magnetizations that form into skyrmions. Recent intensive studies have uncovered many interesting microwave-related phenomena of magnetic skyrmions, and knowledge of their microwave device functions has been accumulated~\cite{Mochizuki12,Mochizuki13,Okamura13,Takeuchi18,Ikka18,Takeuchi19,Koide19}. Moreover, magnetic skyrmions have attracted a great deal of research interest because of their potentials for applications to spin-based electronics. Our proposal paves an alternative path in creating magnetic skyrmions as building blocks of skyrmion-based electronic devices including memories, logic gates, and microwave detectors and provides challenging issues in fundamental science~\cite{Fert13,Tomasello14,Koshibae15,Finocchio16,ZhangX15,Finocchio15}.

It should be mentioned that there have been a lot of important proposals about switching, creation, and excitation of a magnetic vortex or magnetic skyrmion confined in a nanodot using the microwave irradiation so far~\cite{KimSK08a,KimSK08b,ChoiYS10,ZhangB15,LiZX17}. These methods cleverly avoid the difficulty in squeezing a spot of microwave application by restricting the magnetization distribution to a nanometric dot-shaped sample. Our proposal is distinct from these previous methods and has an advantage against them. Specifically, our method can create magnetic skyrmions in a device or sample of arbitrary shape and size. Indeed, the skyrmion-hosting nanodots may be technically useful for microwave detection/generation devices and skyrmion-based magnetoresistive random access memories (skyrmion-MRAM). However, a lot of potential applications of skyrmions such as skyrmion-based race-track memories, magnonic crystals, logic gates, and brain-inspired computation devices require skyrmions hosted in a thin-plate sample large enough to realize moving, alignment and mutual interaction of skyrmions in it. In this sense, our method proposed here may be useful for these kinds of technical applications. 

\section{Acknowledgments}
This work was supported by JSPS KAKENHI (Grants Nos. 17H02924, 16H06345, 19H00864, and 19K21858), Waseda University Grant for Special Research Projects (Project Nos. 2018K-257 and 2019C-253), and JST PRESTO (Grant No. JPMJPR132A).

\end{document}